\documentclass[useAMS,usenatbib,iop]{emulateapj}
\usepackage{natbib}
\usepackage{color,graphicx} 
\usepackage{amsmath}        
\usepackage{amsfonts}       
\usepackage{amssymb}        
\usepackage{wasysym}        
\usepackage{gensymb}        
\usepackage{upgreek}

\shorttitle{A very cool neutron star}
\shortauthors{E.~F.~Keane et al.}

\begin{document}
\title{PSR~J1840$-$1419: A very cool neutron star}

\author{E.~F.~Keane\altaffilmark{1}, M.~A.~McLaughlin\altaffilmark{2},
  M.~Kramer\altaffilmark{1,3}, B.~W.~Stappers\altaffilmark{3},
  C.~G.~Bassa\altaffilmark{3}, M.~B. Purver\altaffilmark{3} \&
  P.~Weltevrede\altaffilmark{3}}

\affil{$^1$ Max Planck Institut f\"{u}r Radioastronomie, Auf dem
  H\"{u}gel 69, 53121 Bonn, Germany. \\$^2$ Department of Physics,
  West Virginia University, Morgantown, WV 26506, USA. \\$^3$
  University of Manchester, Jodrell Bank Centre for Astrophysics,
  School of Physics \& Astronomy, Manchester M13 9PL, UK.}

\begin{abstract}
  We present upper limits on the X-ray emission for three neutron
  stars. For PSR~J1840$-$1419, with a characteristic age of 16.5~Myr,
  we calculate a blackbody temperature upper limit (at 99\%
  confidence) of $kT_{\mathrm{bb}}^{\infty}<24^{+17}_{-10}$~eV, making
  this one of the coolest neutron stars known.  PSRs~J1814$-$1744 and
  J1847$-$0130 are both high magnetic field pulsars, with inferred
  surface dipole magnetic field strengths of $5.5\times10^{13}$ and
  $9.4\times10^{13}$~G, respectively. Our temperature upper limits for
  these stars are $kT_{\mathrm{bb}}^{\infty}<123^{+20}_{-33}$~eV and
  $kT_{\mathrm{bb}}^{\infty}<115^{+16}_{-33}$~eV, showing that these
  high magnetic field pulsars are not significantly hotter than those
  with lower magnetic fields. Finally, we put these limits into
  context by summarizing all temperature measurements and limits for
  rotation-driven neutron stars.
\end{abstract}


\section{Introduction}
Measurements of the temperatures of neutron stars (NSs) are few and
far between. They are often impossible as, for ages $\gtrsim 1$~Myr,
the star will have experienced significant cooling since its
formation~\citep{yp04}. The thermal emission from NSs peaks in soft
X-rays, below $\sim1$~keV, and the luminosities are such that sources
even a few kiloparsecs away are difficult to detect. Currently, a
total of 37 measurements, and 9 upper-limit measurements, of
rotation-driven NS temperatures have been published
\citep{lll05,cdm+10,aaa+10a,skk11,cpks11,zkm+11,mdc11,ggo12,pmt+12}. Of
these, only 11 are for NSs with characteristic ages older than $1$~Myr
(excluding 5 millisecond pulsars for whom we have no meaningful
estimate of age), and only three of those greater than $10$~Myr:
PSRs~B0950$+$08, J2144$-$3933 and J0108$-$1431. The emission from
PSR~B0950$+$08, a $17.5$~Myr pulsar, is non-thermal, with a power-law
spectrum, and an upper limit of $kT_{\mathrm{bb}}^{\infty}<41$~eV, or
$T<0.48$~MK, on the thermal contribution~\citep{bwt+04}. To put this
in context, it is at the level of the lowest measured temperature for
any NS: $kT_{\mathrm{bb}}^{\infty}=43$~eV for
Geminga~\citep{dcm+05}. PSR~J2144$-$3933 is the slowest spinning
pulsar known ($P=8.5$~s) and an order of magnitude older ($272$~Myr)
than PSR~B0950$+$08 so that it is perhaps not surprising that no X-ray
emission has been seen from it, with an upper limit of
$kT_{\mathrm{bb}}^{\infty}<20$~eV~\citep{tmd+11}. \citet{pap+12} have
recently reported the detection of PSR~J0108$-$1431 ($170$~Myr) with
$kT_{\mathrm{bb}}^{\infty}=110$~eV, a source which seems to be
undergoing a heating process.

In a recent re-analysis of the Parkes Multi-beam Pulsar Survey,
PSR~J1840$-$1419 was discovered~\citep{kle+10}. This $6.6$-s pulsar
shows sporadic radio emission with strong single pulses detected, on
average, every $\sim10$ rotation periods, at flux densities up to
$1.7$~Jy. At $16.5$~Myr, PSR~J1840$-$1419 has a very similar
characteristic age to PSR~B0950$+$08, and therefore might be expected
to have somewhat similar X-ray properties. This, plus its relative
proximity at $850$~pc, make PSR~J1840$-$1419 an excellent target for
adding to our knowledge of the X-ray emission and temperature of old
NSs. In this paper we present the result of a \textit{Chandra}
observation of PSR~J1840$-$1419.

One factor which has been suggested to be a major factor in NS cooling
is the magnetic field strength. Several authors have suggested that
NSs with higher magnetic field strengths are hotter than those with
lower field strengths, for the same
age~\citep{gkc+07,zkgl09,oklk10,zkm+11}. Such an effect is
theoretically predicted \citep{yp04} and the work of \citet{gkp04} and
\citet{pmp06} shows that high magnetic fields supress heat
conductivity perpendicular to the field lines, naturally producing an
anisotropic temperature distribution on the stellar surface with small
hot regions at the magnetic poles.
We examine this claim with \textit{XMM-Newton} observations of two
pulsars: J1814$-$1744 and J1847$-$0130, both of which have high
magnetic field strengths, of $5.5\times10^{13}$ and
$9.4\times10^{13}$~G, respectively.

Finally, we present an updated plot of $kT_{\mathrm{bb}}^{\infty}$ vs
characteristic age for all X-ray measurements of, and upper limits on,
rotation-powered NS temperature made thus far, and re-examine the
temperature-magnetic field strength relationship.

\section{Observations \& Results}

\subsection{Calculating Limits}
For a non-detection, one could determine a count rate limit by
following the procedure of \citet{pkc00}: (i) define an aperture on
the image about the position of the source, and add up the counts in
this region, $N$; (ii) define a background region of the same area (or
appropriately scale to the same area) elsewhere on the image and add
up the counts in the background region, $B$. The signal from the
source is $N - B$ and the noise is $\sqrt{N+B}$. The signal-to-noise
ratio is then $(N-B)/\sqrt{N+B}$. To set a `$3-\sigma$ limit' count
rate we can then solve for $N$ in $9(N+B)=(N-B)^2$, where $B$ is
known. We can see that in the $N \gg B$ `photon-rich' case this tends
to the expected Poisson value in the absence of background sky
noise. However, in the `photon-poor' case we clearly can not use this
expression, e.g. for $B = 0.35$, $N=10$ na\"{i}vely suggests a 99.7\%
detection/limit, however we know that the Poisson probability for $10$
counts due to noise is $\approx 10^{-13}\%$; for such a background
rate $3$ counts represents a 99.8\% limit. The observations presented
here are in the photon-poor scenario, so that using the \citet{pkc00}
procedure would result in a much poorer limit than what has truly been
obtained.

To convert count rate limits to flux limits we need either some kind
of absolute scale, e.g. from observing a standard source of known
flux, or a model for our source, e.g. an assumed spectral
form~\citep{msk+03}. Below we calculate one set of limits for a
blackbody model (whence we obtain temperature limits as a function of
the emitting radius), and another set of limits for a non-thermal
model with a power law with negative index of $2.0$.

\subsection{PSR~J1840$-$1419}
PSR~J1840$-$1419 is an old pulsar ($16.5$~Myr) with a $6.6$-s spin
period. Due to its sporadic radio emission it was detected in a search
for single pulses, rather than in a periodicity
search~\citep{kle+10}. Pulsars discovered in this way are often
referred to as ``RRATs''~\citep{km11}. The properties of
PSR~J1840$-$1419, both measured and derived, are given in
Table~\ref{tab:properties}.

On 2011-02-20, we performed a $10$-ks observation of PSR~J1840$-$1419,
using the ACIS-S detector on \textit{Chandra} in the energy range
$0.2-10.0$~keV. We detected only one photon within the $\sim1$~arcsec
($3$ pixel) error circle of the position derived from pulsar
timing~\citep{kkl+11}. Assuming that this 1 count is consistent with
background noise (the background rate being $0.24$~counts/pixel), and
ignoring the Poisson nature of the source, the count rate limit is
$3\times10^{-4}$~counts/s (at $99\%$ confidence). 

\begin{figure*}
  \begin{center}
    \includegraphics[trim = 10mm 0mm 0mm 20mm, clip,scale=0.3,angle=0]{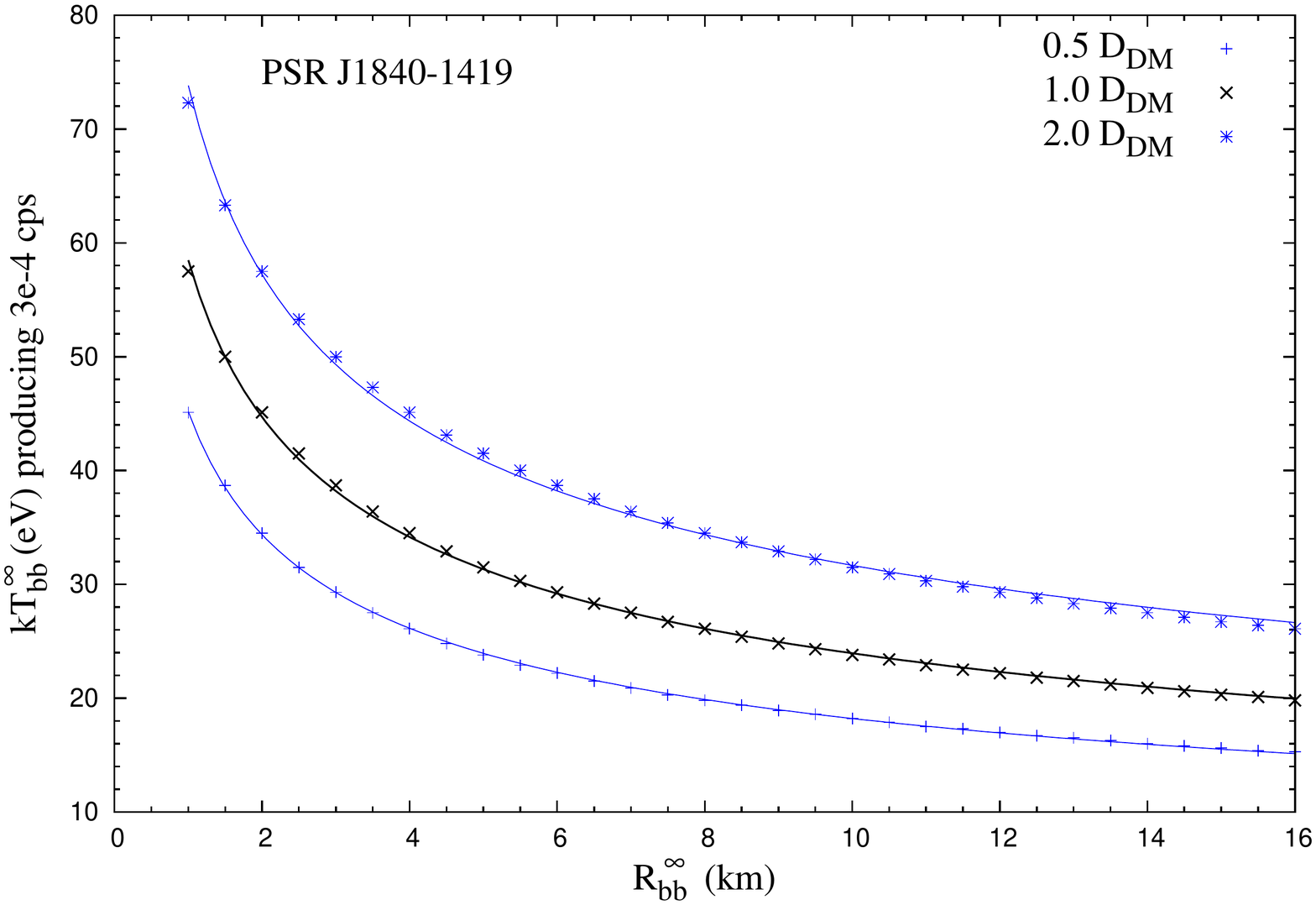}
    \includegraphics[trim = 10mm 0mm 0mm 20mm, clip,scale=0.3,angle=0]{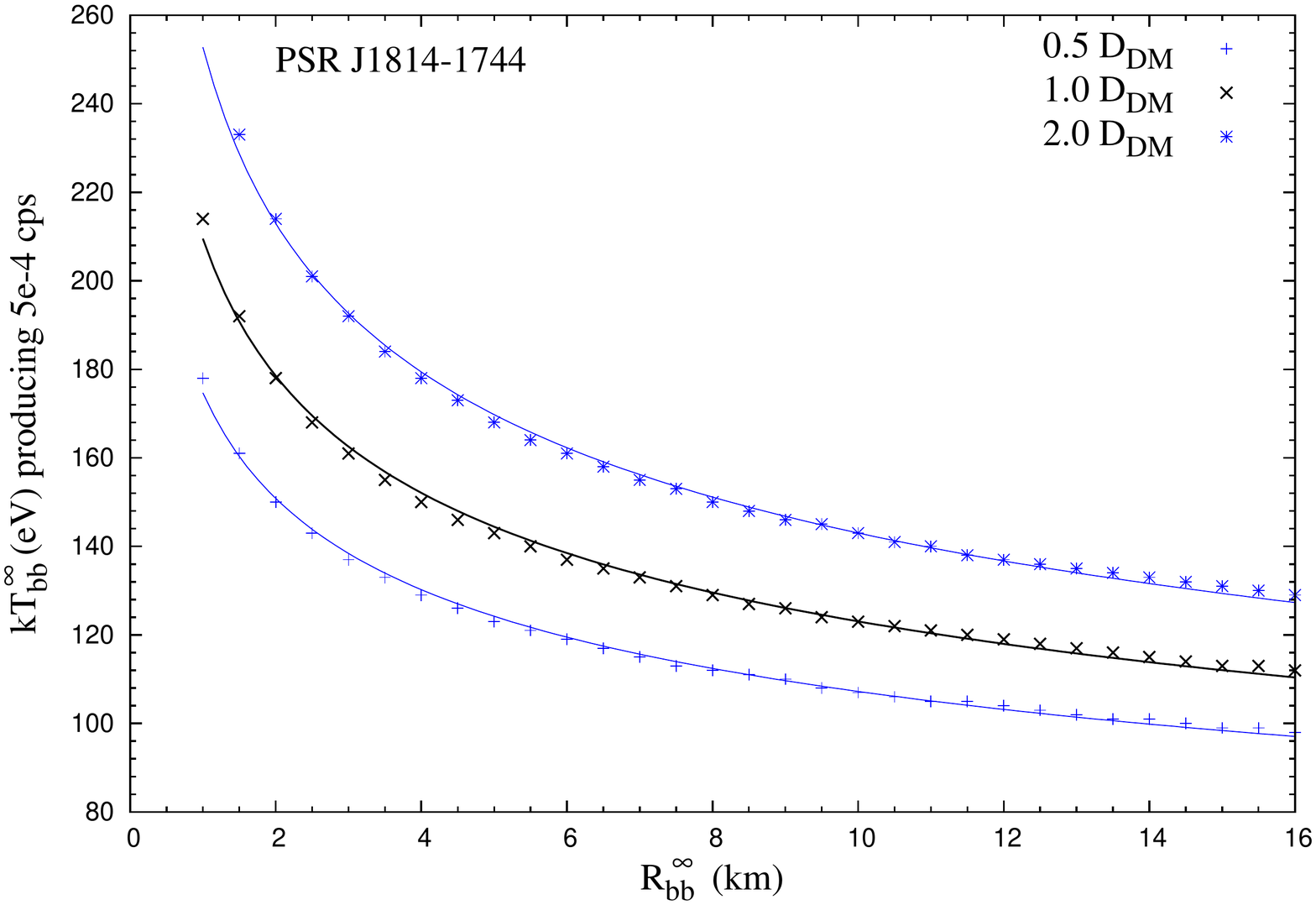}
    \includegraphics[trim = 10mm 0mm 0mm 20mm, clip,scale=0.3,angle=0]{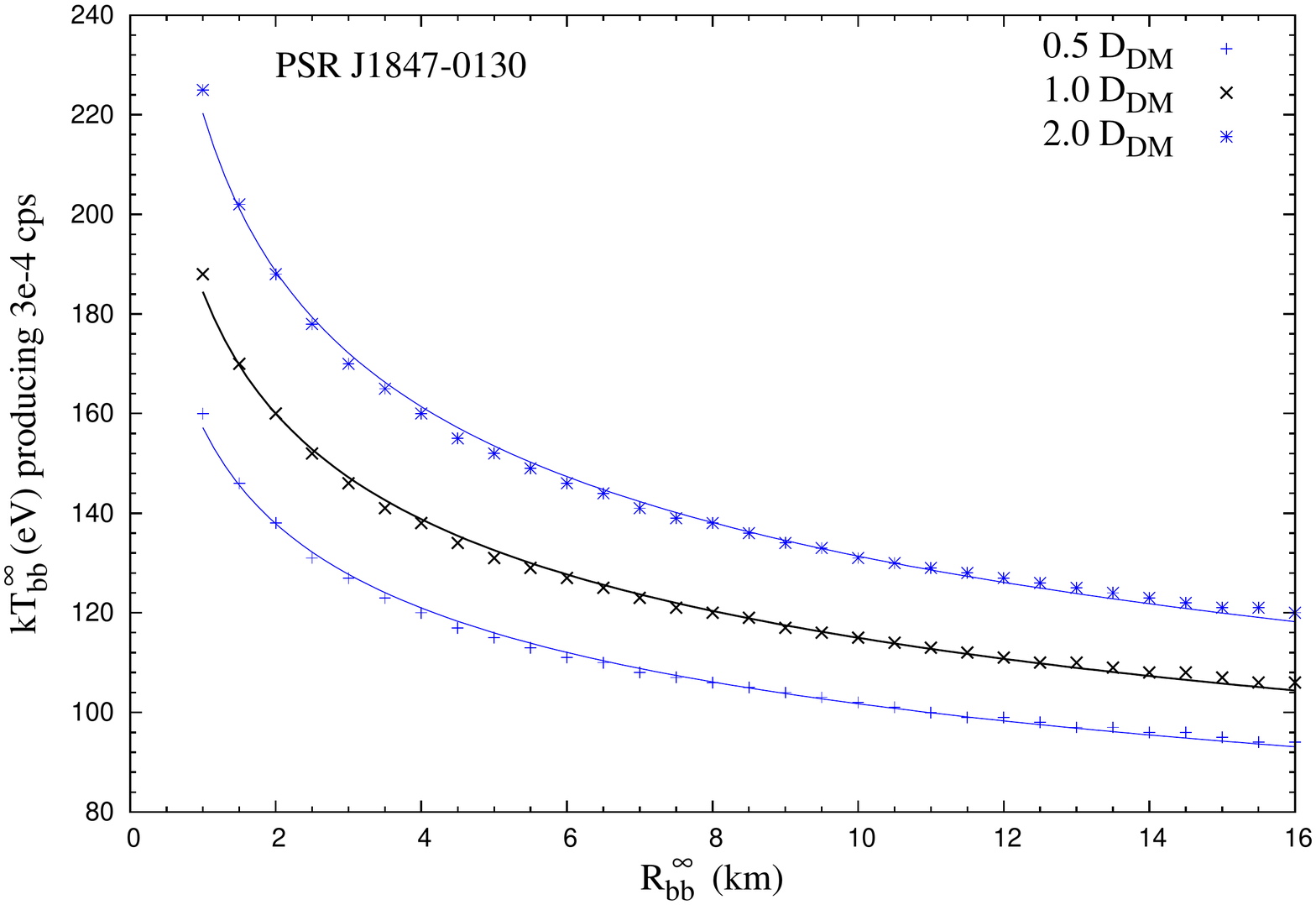}
    \includegraphics[trim = 10mm 0mm 0mm 20mm, clip,scale=0.3,angle=0]{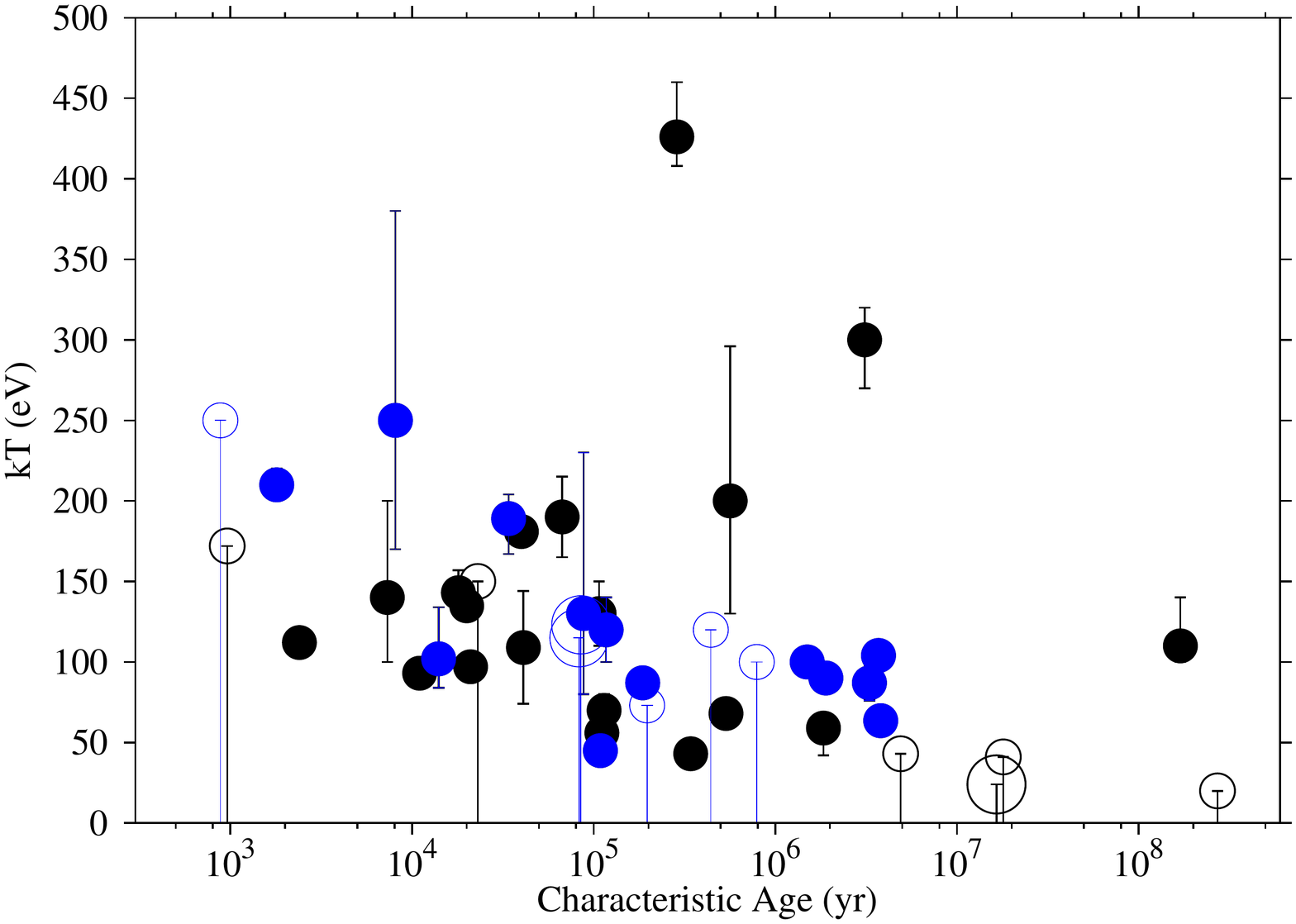}
  \end{center}
  \vspace{-25pt}
  \caption{\small{The first three plots show the temperature limit
      (i.e. which produces the observed count rate limit for an
      assumed blackbody model) as a function of the assumed emitting
      radius, for the \textit{Chandra} observation of PSR~J1840$-$1419
      ($10$~ks) and the \textit{XMM-Newton} observations of
      PSRs~J1814$-$1744 ($6$~ks) and J1847$-$0130 ($17$~ks). The
      points are fit with a simple power-law of the form
      $T_{10}(R_{10})^{\alpha}$. In each case the middle curve is for
      the nominal dispersion-measure-derived distance. The cooler and
      hotter curves are for `worst-case' distance errors of a factor
      of $2$~\citep{cl02}. For clarity of presentation we show here
      only the curves for the nominal $N_{\mathrm{H}}$ values quoted
      in Table~\ref{tab:properties}. The fourth plot shows all known
      temperature values as a function of characteristic age. Sources
      with inferred magnetic field strengths $\geq 10^{13}$~G are
      marked in blue, whereas those with lower values are marked in
      black. Open circles denote upper limits. The three upper limit
      presented in this paper are marked with large circles for
      clarity. Note that the symbols for PSRs~J1814$-$1744 and
      J1847$-$0130 are largely overlapping as their ages and derived
      limits are very similar.}}
  \label{fig:tdist}
\end{figure*}

In the case of a blackbody model this count rate limit implies a
temperature limit of $kT_{\mathrm{bb}}^{\infty} < 24^{+17}_{-10}
\mathrm{eV} (R_{\mathrm{bb}}^{\infty}/10\;\mathrm{km})^{-0.39}$,
where the dependence on $R_{\mathrm{bb}}^{\infty}$ is a purely
empirical fit to a simple power law of the form
$T_{10}R_{10}^{\alpha}$, where $R_{10}$ is the blackbody radius in
units of $10$~km and $T_{10}$ is the temperature when $R_{10}=1$: see
top left panel of Figure~\ref{fig:tdist}. The error bars reflect both
NE2001 `worst case' errors of a factor of two in the
distance~\citep{cl02} as well as factor of two uncertainties in the
neutral hydrgoen column density (see below). The bottom right panel of
Figure~\ref{fig:tdist} shows all measured neutron star temperatures as
a function of age. The PSR~J1840$-$1419 limit is amongst the coolest
of all published limits. The corresponding flux and luminosity limits
are $f_{\mathrm{bb}} < 5.0\times10^{-14}
(R_{\mathrm{bb}}^{\infty}/10\;\mathrm{km})^{1.61}\;
\mathrm{erg}\;\mathrm{s}^{-1}\;\mathrm{cm}^{-2}$ and
$L_{\mathrm{bb}}^{\infty} < 4.3\times10^{30}
(R_{\mathrm{bb}}^{\infty}/10\;\mathrm{km})^{1.61}
\;\mathrm{erg}\;\mathrm{s}^{-1}$, respectively. In the case of the
non-thermal model (power law, with negative index of $2.0$), the flux
and luminosity limits are $f_{\mathrm{nt}} < 3.0\times10^{-15}\;
\mathrm{erg}\;\mathrm{s}^{-1}\;\mathrm{cm}^{-2}$ and
$L_{\mathrm{nt}}^{\infty} < 2.6\times10^{29}
\;\mathrm{erg}\;\mathrm{s}^{-1}$, respectively.

The above estimates use a neutral hydrogen column density of
$N_{\mathrm{H}}=6\times10^{20}\;\mathrm{cm}^{-2}$ which is derived
from the dispersion measure, assuming 10 neutral hydrogen atoms per
free electron~\citep{se98}. Predicted count levels were calculated
using the standard PIMMS (Portable, Interactive Multi-Mission
Simulator) tool\footnote{http://asc.harvard.edu/toolkit/pimms.jsp}.

\subsection{PSRs~J1814$-$1744 \& J1847$-$0130}
We also present the results of two \textit{XMM-Newton} observations
using the PN detector and medium filter in the energy range
$0.15-15.0$~keV. PSR~J1814$-$1744 was observed for $6.1$~ks on
2004-10-21, and PSR~J1847$-$0130 was observed for $17.0$~ks on
2004-09-14. The observations of these young, high-B pulsars also
resulted in non-detections. As above for PSR~J1840$-$1419 we derive
temperature upper limits for these sources of:
$kT_{\mathrm{bb}}^{\infty} < 123^{+20}_{-34} \mathrm{eV}
(R_{\mathrm{bb}}^{\infty}/10\;\mathrm{km})^{-0.23}$, for
PSR~J1814$-$0130, and $kT_{\mathrm{bb}}^{\infty} < 115^{+16}_{-33}
\mathrm{eV} (R_{\mathrm{bb}}^{\infty}/10\;\mathrm{km})^{-0.21}$, for
PSR~~J1847$-$0130 (see Figure~\ref{fig:tdist}). These limits are much
less constraining given the much larger estimated distances ($7.7$ and
$9.8$~kpc respectively). The associated flux and luminosity limits for
the blackbody scenario are: $f_{\mathrm{bb}} < 2.6\times10^{-13}
(R_{\mathrm{bb}}^{\infty}/10\;\mathrm{km})^{1.77}\;
\mathrm{erg}\;\mathrm{s}^{-1}\;\mathrm{cm}^{-2}$ and
$L_{\mathrm{bb}}^{\infty} < 3.0\times10^{33}
(R_{\mathrm{bb}}^{\infty}/10\;\mathrm{km})^{1.77}
\;\mathrm{erg}\;\mathrm{s}^{-1}$ for PSR~J1814$-$1744; and
$f_{\mathrm{bb}} < 3.2\times10^{-13}
(R_{\mathrm{bb}}^{\infty}/10\;\mathrm{km})^{1.79}\;
\mathrm{erg}\;\mathrm{s}^{-1}\;\mathrm{cm}^{-2}$ and
$L_{\mathrm{bb}}^{\infty} < 2.3\times10^{33}
(R_{\mathrm{bb}}^{\infty}/10\;\mathrm{km})^{1.79}
\;\mathrm{erg}\;\mathrm{s}^{-1}$ for PSR~J1847$-$0130.

The non-thermal limits are: $f_{\mathrm{nt}} < 8.9\times10^{-15}\;
\mathrm{erg}\;\mathrm{s}^{-1}\;\mathrm{cm}^{-2}$ and
$L_{\mathrm{nt}}^{\infty} < 1.0\times10^{32}
\;\mathrm{erg}\;\mathrm{s}^{-1}$ for PSR~J1814$-$1744; and:
$f_{\mathrm{nt}} < 5.1\times10^{-15}\;
\mathrm{erg}\;\mathrm{s}^{-1}\;\mathrm{cm}^{-2}$ and
$L_{\mathrm{nt}}^{\infty} < 3.6\times10^{31}
\;\mathrm{erg}\;\mathrm{s}^{-1}$ for PSR~J1847$-$0130.  For
PSR~J1814$-$1744 the value of $N_{\mathrm{H}}$ derived from the
dispersion measure was $\sim50\%$ higher than the maximum Galactic
value for this line of sight, as calculated by the standard COLDEN
tool\footnote{http://asc.harvard.edu/toolkit/colden.jsp}, so the
COLDEN value of $N_{\mathrm{H}}=1.6\times10^{22}\;\mathrm{cm}^{-2}$
was used. For this reason $N_{\mathrm{H}}$ is unlikely to be
underestimated, but we have allowed for the possibility that it may be
overestimated by as much as a factor of two. Likewise, for
PSR~J1847$-$0130, the neutral hydrogen column density was derived to
be $N_{\mathrm{H}}=2.1\times10^{22}\;\mathrm{cm}^{-2}$. This value of
$N_{\mathrm{H}}$ is very close to the maximum value for this line of
sight, so that it too is unlikely to be an overestimate.

\section{Discussion \& Conclusions}\label{sec:conc_disc}
Although this paper reports three null results we deem it to be of the
utmost importance to report such investigations to avoid duplicated
efforts, wasted telescope resources, etc. Table~\ref{tab:properties}
summarises the temperature, flux and luminosity limits for the sources
reported here. From the non-thermal luminosity limits we can determine
upper-limits on the non-thermal X-ray efficiency,
$\eta=L_{\mathrm{nt}}/\dot{E}$, where $\dot{E}$ is the spin-down
energy loss rate. Although the luminosity limits for the
\textit{XMM-Newton} observations are $\sim2$ orders of magnitude
poorer than for the \textit{Chandra} observation of J1840$-$1419, the
$\dot{E}$ values are correspondingly higher, such that for all 3
pulsars the limit turns out to be $\eta \lesssim 0.2$. \citet{pccm02}
determined, for their study of 39 pulsars, that all sources had a
value of $\eta$ less than $\eta_{\mathrm{max}} =
10^{-18.5}(\dot{E}/\mathrm{erg}\;\mathrm{s}^{-1})^{0.48}$. Our $\eta$
limits are all significantly higher than this predicted
value. However, we note this critical $\eta$ value is arrived at upon
considering pulsars studied in the $2-10$~keV energy range, narrower
than the observations reported here.

These observations bring the total number of rotation-powered NSs with
detections of thermal emission, or upper limits thereupon, to 49. In
Figure~\ref{fig:tdist} we show our limits, along with all previous
measurements and limits. \citet{zkm+11} suggest that there is a
``hint'' that the high magnetic field pulsars are hotter than the low
magnetic field pulsars. To quantify this we first ignored the upper
limits and divided the remaining data into `low' and `high'
magnetic-field-strength sources, below and above an arbitrarily chosen
field strength value of $10^{13}$~G. We also excluded three `low'
magnetic-field-strength sources as their fitted blackbody radii are so
much smaller than those of the rest of the sources (at $33$, $43$ and
$120$~m) so that their thermal emission is thought to be due to some
heating process~\citep{mpg08}. Comparing the two distributions with a
Kolmogorov-Smirnov (K-S) test we find a K-S statistic of $D=0.33$, and
thus a probability of $0.31$ that these two distributions are the
same. There is therefore no evidence that the high magnetic field
pulsars are hotter than those with lower magnetic fields. However we
note the caveat that, in addition to the age estimates being uncertain
(as mentioned above), the magnetic field estimates are uncertain, and
should only be considered accurate to within an order of
magnitude. Knowledge of the inclination angles between the magnetic
and rotation axes would be needed to estimate the magnetic field
strengths more accurately (e.g. as in \cite{spi06}).

\section*{Acknowledgments}
This work has been supported by NASA CXO guest observer support grant
GO1-12059X, and made use of software provided by the Chandra X-ray
Center. EFK thanks the FSM for support, K. J. Lee for useful
discussions on statistics, R. P. Eatough for useful comments on the
manuscript and the anonymous referee for providing valuable input that
improved the quality of this paper. MAM is supported by the Research
Corporation for Scientific Advancement.

\begin{deluxetable*}{lccc}
  \tablecaption{Measured and derived properties, and limits for 3 pulsars.\label{tab:properties}}
  \tablecolumns{4}
  \tablehead
  {
    \colhead{Quantity} & 
    \colhead{J1840$-$1419} &
    \colhead{J1814$-$1744} &
    \colhead{J1847$-$0130} \\
  }
      
  \startdata
  $P$ (s) & $6.6$ & $4.0$ & $6.7$ \\
  $\dot{P}$ (fs/s) & 6.3 & 745 & 1270 \\
  $\tau_{\mathrm{c}}$ (kyr) & 16500 & 85 & 83 \\
  $B$ ($10^{12}$ G) & 6.5 & 55.1 & 93.6 \\
  $\dot{E}$ ($10^{30}\;\mathrm{erg}\;\mathrm{s}^{-1}$) & 1.0 & 468 & 167 \\
  $DM$ ($\mathrm{cm}^{-3}\;\mathrm{pc}$) & 19 & 792 & 667 \\
  $D$ (kpc) & 0.9 & 9.8 & 7.7 \\
  Telescope & Chandra & XMM & XMM \\
  Instrument & ACIS-S & PN & PN \\
  Date & 2011-02-20 & 2004-10-21 & 2004-09-14 \\
  Energy Range (keV) & $0.2-10$ & $0.15-15$ & $0.15-15$ \\
  $T_{\mathrm{obs}}$ (ks) & 10.0 & 6.1 & 17.0 \\
  $N_{\mathrm{H}}$ ($10^{20}\;\mathrm{cm}^{-2}$) & 6 & 157 & 205 \\
  $kT_{\mathrm{bb}}^{\infty}$ (eV)$\dagger$ & $<24^{+17}_{-10}$ & $<123^{+20}_{-34}$ & $<115^{+16}_{-33}$ \\
  $T_{\mathrm{bb}}^{\infty}$ (MK)$\dagger$ & $<0.28^{+0.19}_{-0.12}$ & $<1.42^{+0.22}_{-0.39}$ & $<1.33^{+0.19}_{-0.38}$ \\
  $f_{\mathrm{bb}}$ ($10^{-14}\;\mathrm{erg}\;\mathrm{cm}^{-2}\;\mathrm{s}^{-1}$)$\ddagger$ & $<5.0$ & $<6.1$ & $<8.9$ \\ 
  $L_{\mathrm{bb}}$ ($10^{30}\;\mathrm{erg}\;\mathrm{s}^{-1}$)$\ddagger$ & $<4.3$ & $<692$ & $<640$ \\
  $f_{\mathrm{nt}}$ ($10^{15}\;\mathrm{erg}\;\mathrm{cm}^{-2}\;\mathrm{s}^{-1}$) & $<3.0$ & $<8.9$ & $<5.1$ \\ 
  $L_{\mathrm{nt}}$ ($10^{30}\;\mathrm{erg}\;\mathrm{s}^{-1}$) & $<0.26$ & $<102$ & $<36$ \\
  $\eta=L_{\mathrm{nt}}/\dot{E}$ & $<0.26$ & $<0.22$ & $<0.22$ \\
  \enddata

  \tablecomments{The measured and derived properties for the three
    pulsars discussed in this paper. The spin period, $P$, and its
    derivative, $\dot{P}$, are derived from pulsar timing techniques,
    and $\tau$, $B$ and $\dot{E}$, the characteristic age, magnetic
    field strength and spin-down energy respectively, are inferred
    from these~\citep{kkl+11}. The distance, $D$, is derived from the
    measured $DM$ using the NE2001 model for the Galactic free
    electron distribution~\citep{cl02}. The $99\%$-confidence
    temperature limits are derived as outlined in the
    text. Non-thermal flux limits are given for a power law with
    negative index of $2.0$. $\dagger$ these limits are for an
    emitting radius of $10$~km, and scale as
    $(R_{bb}^{\infty}/10\;\mathrm{km})^{\alpha}$, where $\alpha$
    equals $-0.39$, $-0.23$ and $-0.21$, respectively for the three
    sources (see main text). $\ddagger$ The blackbody flux and
    luminosity limits similarly scale as
    $(R_{bb}^{\infty}/10\;\mathrm{km})^{2-\alpha}$.}

\end{deluxetable*}

\newpage
\bibliography{journals,journals_apj,psrrefs,modrefs,crossrefs}
\bibliographystyle{mnras}

\end{document}